\documentclass{ws-procs961x669}            
\usepackage{xcolor}
\usepackage{float}
\usepackage{tikz}
\usepackage{hyperref}
\definecolor{purple}{rgb}{1,0,1} 

\newcommand{\orcidicon}{%
	\begin{tikzpicture}
	\draw[lime, fill=lime] (0,0) 
		circle [radius=0.16] 
		node[white] {{\fontfamily{qag}\selectfont \tiny ID}};
	\draw[white, fill=white] (-0.0625,0.095) 
		circle [radius=0.007];
	\end{tikzpicture}
	\hspace{-5mm}
}
\newcommand\orcidAlex{{\href{https://orcid.org/0000-0002-1763-3563}{\orcidicon}}}

\newcommand{\dd}{\mathrm{d}}

\begin{document}
\title{From black-bounce to traversable wormhole, and beyond}

\author{Alex Simpson\!\orcidAlex\!}

\address{School of Mathematics and Statistics, Victoria University of Wellington, PO Box 600, Wellington 6140, New Zealand.\\
E-mail: alex.simpson@sms.vuw.ac.nz}

\begin{abstract}
Key results from the literature pertaining to a class of nonsingular black hole mimickers are explored. The family of candidate spacetimes is for now labelled the `black-bounce' family, stemming from the original so-called `Simpson--Visser' spacetime in static spherical symmetry. All model geometries are analysed through the lens of standard general relativity, are globally free from curvature singularities, pass all weak-field observational tests, and smoothly interpolate between regular black holes and traversable wormholes. The discourse is segregated along geometrical lines, with candidate spacetimes each belonging to one of: static spherical symmetry, spherical symmetry with dynamics, and stationary axisymmetry.
\end{abstract}


\bodymatter


\section{Introduction}

In classical general relativity (GR), solutions to the Einstein equations representing black holes typically contain curvature singularities at their cores. These singularities occur at a distance scale that only a complete theory of quantum gravity could adequately describe. In the absence of such a theory, or at least in the absence of one that has any phenomenological verification, one desirable approach is to appeal to the recent advances made in observational and gravitational wave astronomy. In view of the propagation of gravitational waves emanating from an astrophysical source being directly observed in the recent LIGO/VIRGO merger events~\cite{ligo-detection-papers, grav-wave-observations-wiki}, it is well-motivated to explore mathematically tractable candidate spacetimes, which are curvature-singularity-free alternatives to classical black holes, through the lens of standard GR. The community hopes that LIGO/VIRGO (or more likely LISA~\cite{LISA}) will eventually be able to provide phenomenological evidence that allows us to delineate between specific candidate spacetimes based on their astrophysical accuracy. Theoretical physicists would then have experimentally informed clues as to which modifications to the Einstein equations are most desirable in working towards a so-called `theory of everything' (or, indeed, whether an entirely different framework is required). Discussion regarding the extraction of astrophysical observables for nonsingular sample spacetimes in several frameworks can be found in references~\citenum{Eiroa:2010, Flachi:2012, Abdujabbarov:2016, Carballo-Rubio:2018, Carballo-Rubio:2019a, Carballo-Rubio:2019b, Dai:2019, Simonetti:2020, Berry:2020, Carballo-Rubio:2021, Bronnikov:2021, Churilova:2021, Bambi:2021, Simpson:2021biv}.

One such family of candidate geometries which models alternatives to classical black holes is the family of `black-bounce' spacetimes~\cite{Simpson:2018, Lobo:2020a, Simpson:2019, Lobo:2020b, Mazza:2021, Franzin:2021}. All of these black hole mimickers are globally free from curvature singularities, pass all weak-field observational tests of standard GR, and belong either to the class of regular black holes or traversable wormholes. By `regular black hole', one means in the sense of Bardeen~\cite{Bardeen:1968}, with regularity defined \emph{via} enforcing global finiteness on nonzero curvature tensor components and Riemann curvature invariants. Regular black holes have a well-established lineage in the historical literature~\cite{Bardeen:1968, Ayon-Beato:2000, Hayward:2005, Bambi:2013, Frolov:2014, Neves:2014, Fan:2016}. By `traversable wormhole', one means in the sense of Morris and Thorne~\cite{Morris:1988a}; a horizon-free geometry with a centralised throat connecting two asymptotically Minkowski regions of spacetime and satisfying the `flare-out' condition for the area function: $A''(r_{throat})>0$. Various intriguing wormhole spacetimes have been developed and explored in references~\citenum{Morris:1988a, Morris:1988b, Visser:1989a, Visser:1989b}.

For exposition, it is prudent to separate the discourse surrounding the family of black-bounce spacetimes along geometrical lines. Broadly speaking, there are three relevant geometrical categories: static spherical symmetry, spherical symmetry with dynamics, and stationary axisymmetry (this also \emph{somewhat} follows the chronological development of the literature). Unless otherwise stated, all candidate spacetimes discussed have metric signature $(-,+,+,+)$ outside any would-be horizons.

\section{Static spherical symmetry}

\subsection{Simpson--Visser}

The so-called `Simpson--Visser' (SV) spacetime, initially presented in reference~\citenum{Simpson:2018}, is represented by the line element 
\begin{equation}\label{SV}
    \dd s^{2} = -\left(1-\frac{2m}{\sqrt{r^{2}+\ell^{2}}}\right)\,\dd t^{2} + \frac{\dd r^{2}}{\left(1-\frac{2m}{\sqrt{r^{2}+\ell^{2}}}\right)}+\left(r^{2}+\ell^{2}\right)\,\dd\Omega^{2}_{2} \ .
\end{equation}
It should be noted that the parameter `$\ell$' was in fact labelled `$a$' in the original article~\cite{Simpson:2018}. This minor alteration is performed for consistency with the remainder of the discourse herein, as when in the axisymmetric environment of \S~\ref{S:axi}, it is prudent to use $\ell$ in order to avoid confusion with the spin parameter from Kerr spacetime, $a$.

\begin{figure}[!htb]
\begin{center}
\includegraphics[scale=0.30]{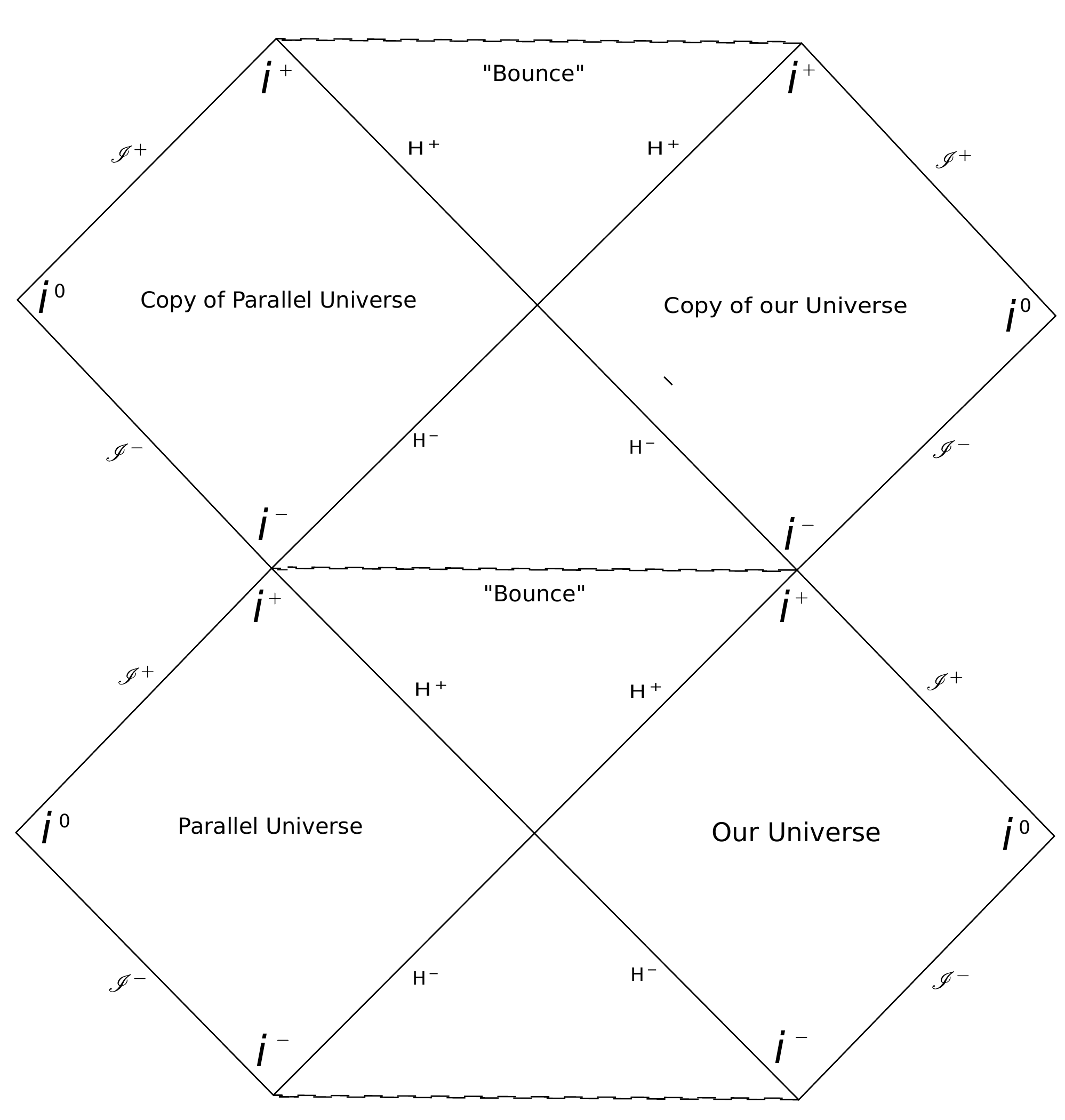}\qquad
\end{center}
{\caption{{Carter--Penrose diagram for the maximally extended spacetime when $\vert\ell\vert\in(0,2m)$. In this example one `bounces' through the $r=0$  hypersurface in each black hole region into a future copy of the universe \emph{ad infinitum}.}}\label{F:bounce-1}}
\end{figure}

\begin{figure}[!htb]
\vspace{-0.5cm}
\begin{center}
\includegraphics[scale=0.30]{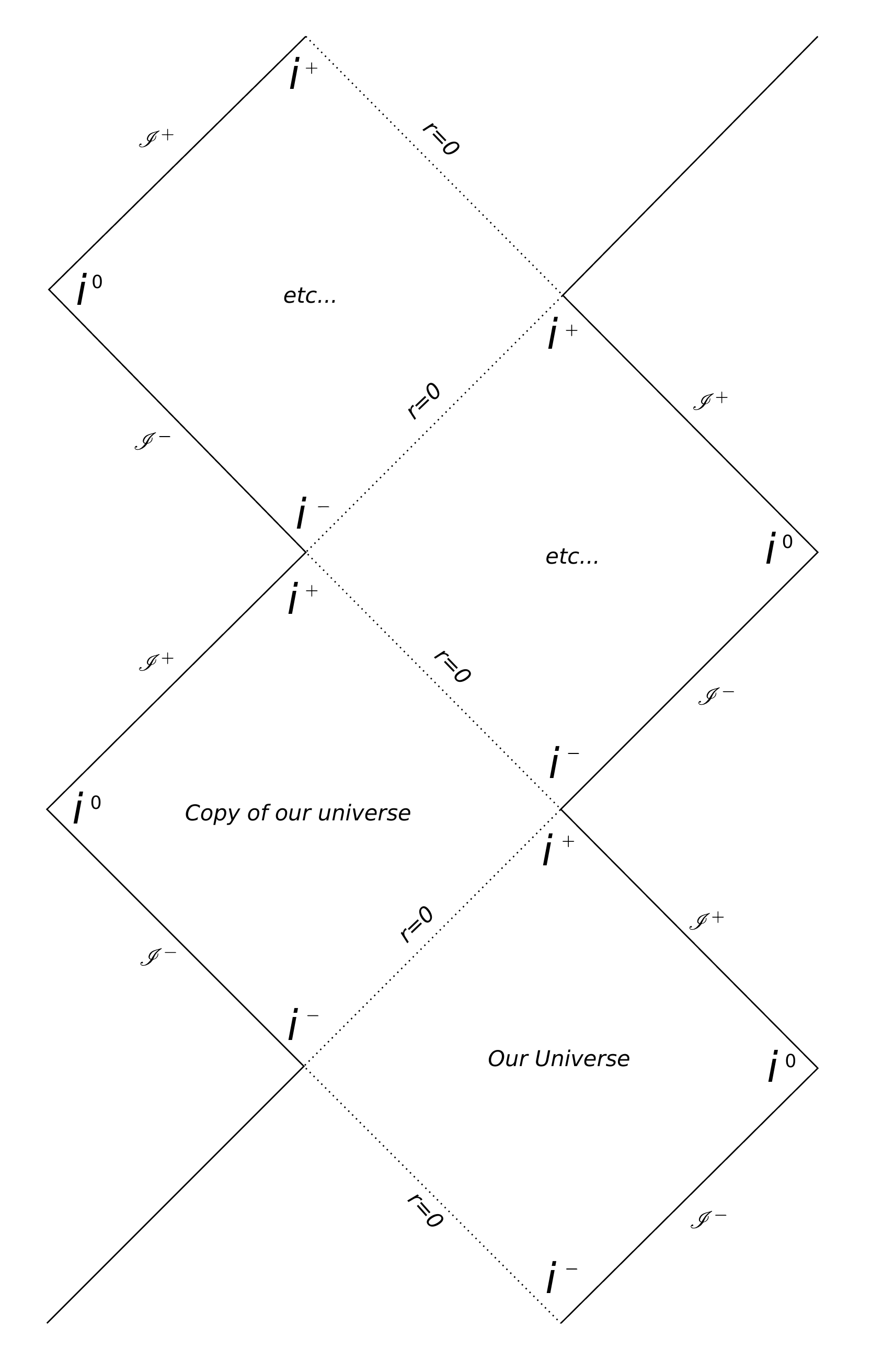}\qquad
\end{center}
{\caption{{Carter--Penrose diagram for the maximally extended spacetime in the case when $\vert\ell\vert=2m$. In this example one has a one-way wormhole geometry with an extremal null throat}.}
\label{F:null-bounce-1}}
\end{figure}

The original motivation for the construction of the metric specified by Eq.~(\ref{SV}) was to minimally modify the Schwarzschild solution in common curvature coordinates such that the resulting candidate spacetime was globally nonsingular. It was considered that the most straightforward way to achieve this was to introduce a new scalar parameter to the line element in a tightly controlled manner; this is $\ell$ in Eq.~(\ref{SV}). By minimally modifying Schwarzschild, it was hoped the result would have a high degree of mathematical tractability. When viewed as a modification of Schwarzschild, there are the following two alterations:
\begin{itemize}
    \item $m \mapsto m(r) = \frac{mr}{\sqrt{r^{2}+\ell^{2}}}$ ;
    \item The coefficient of $\dd\Omega^{2}_{2}$ is modified from $r^2\rightarrow r^{2}+\ell^{2}$ .
\end{itemize}
Analysis of the nonzero curvature tensor components and Riemann curvature invariants concludes that for $\vert\ell\vert>0$, the resulting candidate spacetime is globally regular. It was also noticed that Eq.~(\ref{SV}) has very neat limiting behaviour. In the limit as $m\rightarrow0$, one obtains
    \begin{equation}
        \dd s^{2} = -\dd t^{2} + \dd r^{2} + \left(r^{2}+\ell^{2}\right)\,\dd\Omega^{2}_{2} \ .
    \end{equation}
    This is precisely the two-way traversable wormhole solution as presented in Morris and Thorne's aforementioned seminal paper~\cite{Morris:1988a} (and arguably the most straightforward of all traversable wormhole geometries). In the limit as $\ell\rightarrow0$, Eq.~(\ref{SV}) becomes the Schwarzschild solution in the usual curvature coordinates.
The newly introduced scalar parameter $\ell$ can hence be viewed as quantifying the extent of the deviation away from Schwarzschild, and it invokes a rich, $\ell$-dependent horizon structure. The causal structure is characterised by
\begin{equation}
    r_{H} = \sqrt{(2m)^2-\ell^2} \ ,
\end{equation}
and the candidate spacetime neatly interpolates between the following qualitatively different geometries:
\begin{itemize}
    \item $\ell=0$ corresponds to the Schwarzschild black hole;
    \item $\vert\ell\vert\in\left(0, 2m\right)$ corresponds to a regular black hole in the sense of Bardeen;
    \item $\vert\ell\vert=2m$ corresponds to a one-way wormhole with an extremal null throat;
    \item $\vert\ell\vert>2m$ corresponds to a two-way traversable wormhole geometry in the canonical sense of Morris and Thorne.
\end{itemize}
The Carter--Penrose diagrams for the two lesser known cases are worth closer examination; this is when $\vert\ell\vert\in(0,2m)$ (see Fig.~\ref{F:bounce-1}), and when $\vert\ell\vert=2m$ (see Fig.~\ref{F:null-bounce-1}).

%
Analysing the satisfaction/violation of the standard point-wise energy conditions of GR, one concludes that when $\vert\ell\vert>0$, the radial `null energy condition' (NEC) is manifestly violated in the region $\sqrt{r^2+\ell^2}>2m$. This is \emph{outside} any would-be horizons. In the context of static spherical symmetry, this is sufficient to conclude that all of the standard point-wise energy conditions shall be similarly violated. If horizons are present, then the geometry has surface gravity
\begin{equation}
    \kappa = \frac{\sqrt{(2m)^2-\ell^2}}{8m^2} = \kappa_{Sch.}\,\sqrt{1-\frac{\ell^2}{(2m)^2}} \ ,
\end{equation}
and hence associated Hawking temperature
\begin{equation}
    T_{H} = \frac{\hbar\sqrt{(2m)^2-\ell^2}}{16\pi\,k_{B}\,m^2} = T_{H,Sch.}\,\sqrt{1-\frac{\ell^2}{(2m)^2}} \ .
\end{equation}
SV spacetime is amenable to the extraction of astrophysical observables, and the coordinate locations of the photon sphere for null orbits and the ISCO for timelike orbits are straightforward, given by
\begin{equation}
    r_{\gamma} = \sqrt{(3m)^2-\ell^2} \ ; \qquad r_{ISCO} = \sqrt{(6m)^2-\ell^2} \ .
\end{equation}
Further research analysing SV spacetime has been performed in a plethora of other papers; please see reference~\citenum{SVinspirepage} for details. These analyses include discussion of the quasi-normal modes and associated ringdown~\cite{Bronnikov:2019, Churilova:2019}, calculations pertaining to shadows and gravitational lensing effects~\cite{Islam:2021, Tsukamoto:2021, Guerrero:2021, Lima:2021, Tsukamoto:2020, Nascimento:2020}, as well as discourse surrounding precession phenomena~\cite{Zhou:2020}.

%
\subsection{Black-bounce Reissner--Nordstr\"{o}m}\label{Ss:bbRN}

The extension of SV spacetime which represents the black-bounce analog to Reissner--Nordstr\"{o}m (RN) spacetime was constructed and analysed in reference~\citenum{Franzin:2021}. There is a simple regularisation procedure which can be applied to any spherically symmetric or axisymmetric geometry in possession of a curvature singularity at $r=0$ in the standard $(t,r,\theta,\phi)$ curvature coordinates. The procedure outputs a candidate geometry which is globally nonsingular, and maintains the manifest symmetries. It is as follows:
\begin{itemize}
    \item Leave the object $\dd r$ in the line element undisturbed;
    \item Whenever the metric components $g_{\mu\nu}$ have an explicit $r$-dependence, replace the $r$-coordinate with $\sqrt{r^2+\ell^2}$, where $\ell$ is some length scale, performing the same holistic role as the $\ell$ parameter in SV spacetime (such parameters are often identified with the Planck scale; $\ell\sim m_{p}$).
\end{itemize}
It should be emphasised that this procedure is \emph{not} a coordinate transformation. Application of this procedure to the usual RN spacetime in standard curvature coordinates yields the following line element, the `black-bounce Reissner--Nordstr\"{o}m' (bbRN) spacetime:
\begin{equation}
    \dd s^2 = -\left(1-\frac{2m}{\sqrt{r^2+\ell^2}}+\frac{Q^2}{r^2+\ell^2}\right)\,\dd t^2 + \frac{\dd r^2}{1-\frac{2m}{\sqrt{r^2+\ell^2}}+\frac{Q^2}{r^2+\ell^2}} + (r^2+\ell^2)\,\dd\Omega^2_2 \ .
\end{equation}
For $\vert\ell\vert>0$, this spacetime neatly interpolates between nonsingular electrovac black holes and traversable wormholes in standard GR, depending on the value of the charge parameter $Q$ and the `bounce' parameter $\ell$. When compared with standard RN spacetime, the domain for the $r$ coordinate is extended from $r\in[0,+\infty)$ to $r\in(-\infty,+\infty)$. The presence of the additional charge term invokes a richer causal structure than that of SV spacetime. Horizons are characterised by
\begin{equation}
    r_{H} = S_{1}\sqrt{(m+S_{2}\sqrt{m^2-Q^2})^2-\ell^2} \ .
\end{equation}
Here $S_1=S_2=\pm 1$, with $S_1$ fixing which universe one is in, and $S_2$ determining whether it is an outer ($S_2=+1$) or inner ($S_2=-1$) horizon. There are several qualitatively different geometries:
\begin{itemize}
    \item $\vert Q\vert>m$: There are no horizons, and the geometry models a two-way traversable wormhole in the sense of Morris and Thorne;
    \item $\vert Q\vert\leq m$ but $\vert\ell\vert>m\pm\sqrt{m^2-Q^2}$: first the inner, and then the outer horizons vanish, leaving a traversable wormhole;
    \item $\vert Q\vert=m$ and $\vert\ell\vert\leq m$: one finds extremal horizons at $r_{H} = \pm\sqrt{m^2-\ell^2}$;
    \item $\vert Q\vert\leq m$ and $\vert\ell\vert< m - \sqrt{m^2-Q^2}$: one observes the `standard' causal structure of both an inner and outer horizon present in each universe.
\end{itemize}
If horizons are present, the surface gravity of the outer horizon is given by
\begin{equation}
    \kappa = \frac{\sqrt{(m+\sqrt{m^2-Q^2})^2-\ell^2}\sqrt{m^2-Q^2}}{(m+\sqrt{m^2-Q^2})^3} = \kappa_{\text{RN}}\sqrt{\frac{r_{H}^2}{r_{H}^2+\ell^2}} \ .
\end{equation}
The following astrophysical observables are amenable to extraction; firstly the location of the photon sphere
\begin{equation}
    r_{\gamma} = \sqrt{\frac{m}{2}(9m+3\sqrt{9m^2-8Q^2})-2Q^2-\ell^2} \ ,
\end{equation}
and also the location of the ISCO for timelike particles
\begin{eqnarray}
    r_{ISCO} &=& \frac{\sqrt{9m^4Q^4-6m^2Q^2(A^2+2Am^2+4m^4)+(A^2+2Am^2+4m^4)^2-A^2\ell^2m^2}}{mA} \ , \nonumber \\
    && \nonumber \\
    A &=& \left[2m^2Q^4+m^2(\pm B-9m^2)Q^2+8m^6\right]^{\frac{1}{3}} \ , \quad B = \sqrt{4Q^4-9m^2Q^2+5m^4} \ . \nonumber \\
    &&
\end{eqnarray}
In standard GR, the stress-energy tensor takes the following form outside the outer horizon (or inside the inner horizon):
\begin{equation}
    \frac{1}{8\pi}G^{\hat{\mu}}{}_{\hat{\nu}} = T^{\hat{\mu}}{}_{\hat{\nu}} = \left[T_{\text{bb}}\right]^{\hat{\mu}}{}_{\hat{\nu}} + \left[T_{Q}\right]^{\hat{\mu}}{}_{\hat{\nu}} = \mbox{diag}(-\rho, p_{r}, p_{t}, p_{t}) \ ,
\end{equation}
whilst in between the two horizons one has
\begin{equation}
    \frac{1}{8\pi}G^{\hat{\mu}}{}_{\hat{\nu}} = T^{\hat{\mu}}{}_{\hat{\nu}} = \left[T_{\text{bb}}\right]^{\hat{\mu}}{}_{\hat{\nu}} + \left[T_{Q}\right]^{\hat{\mu}}{}_{\hat{\nu}} = \mbox{diag}(p_{r}, -\rho, p_{t}, p_{t}) \ ,
\end{equation}
where $\left[T_{\text{bb}}\right]^{\hat{\mu}}{}_{\hat{\nu}}$ is the gravitational stress-energy tensor (identical to that of SV spacetime), and $\left[T_{Q}\right]^{\hat{\mu}}{}_{\hat{\nu}}$ is the charge-dependent contribution, corresponding to the electromagnetic stress-energy tensor. Isolating the charge-dependent components, it is straightforward to make the following decomposition
\begin{eqnarray}
    \left[T_{Q}\right]^{\hat{\mu}}{}_{\hat{\nu}} &=& \frac{Q^2r^2}{8\pi(r^2+\ell^2)^3}\left[\mbox{diag}(-1,-1,1,1)+\mbox{diag}(\frac{2\ell^2}{r^2},0,0,0)\right] \nonumber \\
    && \nonumber \\
    &=& \left[T_{\text{Maxwell}}\right]^{\hat{\mu}}{}_{\hat{\nu}} + \Xi\, V^{\hat{\mu}}V_{\hat{\nu}} \ ,
\end{eqnarray}
where $\left[T_{\text{Maxwell}}\right]^{\hat{\mu}}{}_{\hat{\nu}}$ is the usual stress-energy associated with standard Maxwell electromagnetism, and $\Xi\,V^{\hat{\mu}}V_{\hat{\nu}}$ is interpreted as the stress-energy of ``charged dust", with $V^{\hat{\mu}}$ being the normalised time-translation Killing vector. The density of the charged dust, $\Xi$, is given by
\begin{equation}
    \Xi = -\frac{1}{4\pi}\frac{Q^2\ell^2}{(r^2+\ell^2)^3} \ ,
\end{equation}
while the electric field strength is determined \emph{via}
\begin{eqnarray}\label{ERN}
    \left[T_{Q}\right]^{\hat{t}}{}_{\hat{t}} &=& -\rho_{\text{em}} = \left[T_{\text{Maxwell}}\right]^{\hat{t}}{}_{\hat{t}}-\Xi = -\frac{1}{8\pi}E^2-\Xi \ ; \nonumber \\
    && \nonumber \\
    \Longrightarrow \quad E &=& \frac{Qr}{(r^2+\ell^2)^{\frac{3}{2}}} = E_{\text{RN}}\left[\frac{r^3}{(r^2+\ell^2)^{\frac{3}{2}}}\right] \ ,
\end{eqnarray}
where $E_{\text{RN}}$ is the electric field strength for a usual RN black hole. The electromagnetic potential is readily obtained \emph{via} integration of Eq.~(\ref{ERN}),
\begin{equation}
    A_{\mu} = (\Phi_{\text{em}}(r),0,0,0) = -\frac{Q}{\sqrt{r^2+\ell^2}}(1,0,0,0) \ ,
\end{equation}
which is simply the RN electromagnetic potential under the mapping $r\rightarrow\sqrt{r^2+\ell^2}$. All told, one has:
\begin{eqnarray}
    \left[T_{Q}\right]^{\hat{\mu}}{}_{\hat{\nu}} &=& \frac{1}{4\pi}\left[-F^{\hat{\mu}}{}_{\hat{\alpha}}F^{\hat{\alpha}}{}_{\hat{\nu}}-\frac{1}{4}\delta^{\hat{\mu}}{}_{\hat{\nu}}F^2\right] - \frac{1}{4\pi}\frac{Q^2\ell^2}{(r^2+\ell^2)^3}\,V^{\hat{\mu}}V_{\hat{\nu}} \ , \nonumber \\
    && \nonumber \\
    \mbox{with} \quad F_{\hat{\mu}\hat{\nu}} &=& \nabla_{\hat{\mu}}A_{\hat{\nu}} - \nabla_{\hat{\nu}}A_{\hat{\mu}} \ .
\end{eqnarray}
Consequently, in standard GR the bbRN spacetime possesses stress-energy components which are interpreted as everywhere-SV in the gravitational sector, coupled to standard Maxwell's electromagnetism in the presence of charged dust in the electromagnetic sector.

%
\subsection{Generalised extensions to Simpson--Visser}\label{Ss:NBB}

Further extensions to the SV line element in the context of static spherical symmetry were explored in reference~\citenum{Lobo:2020a}, with some useful general theorems also being presented. The first of these theorems is general to all static spacetimes, and is concerned with expediting the `usual' test for curvature-regularity; examination of the finiteness of all nonzero components of the Riemann curvature tensor in an orthonormal basis. It is as follows, with the proof provided in reference~\citenum{Lobo:2020a}:
\begin{theorem}\label{T21}
    For any static spacetime, in the strictly static region, the Kretschmann scalar is positive semi-definite, being a sum of squares which involves all of the nonzero components $R^{\hat{\mu}\hat{\nu}}{}_{\hat{\alpha}\hat{\beta}}$. Then if this scalar is finite, all of the orthonormal components of the Riemann curvature tensor must also be finite. Consequently, for static candidate spacetimes, confirmation of the global finiteness of the Kretschmann scalar is sufficient to conclude as to curvature-regularity in the sense of Bardeen.
\end{theorem}
It is by now well-known that the most general static spherically symmetric line element can always locally be put into the following form, typically known as ``Buchdahl" coordinates~\cite{Delgaty:1998, Finch:1998, Boonserm:2007a, Boonserm:2007b, Semiz:2020} (for \S~\ref{Ss:NBB} only, the metric signature $(+,-,-,-)$ is adopted outside horizons for consistency with the discourse in reference~\citenum{Lobo:2020a})
\begin{equation}\label{Buch}
    \dd s^2 = f(r)\,\dd t^2 - \frac{\dd r^2}{f(r)} - \Sigma^2(r)\,\dd\Omega^2_2 \ .
\end{equation}
Now constrained to Buchdahl coordinates, it is straightforward to establish the following list of sufficient conditions for curvature regularity in the sense of Bardeen (note these constraints exist as an independent result from Theorem~\ref{T21}; which of the techniques is preferable/applicable should be determined from context):
\begin{itemize}
    \item $\Sigma(r)\neq0$ globally;
    \item $\Sigma'(r)$ and $\Sigma''(r)$ must be globally finite;
    \item $f(r)$, $f'(r)$, and $f''(r)$ must be globally finite.
\end{itemize}
With respect to Buchdahl coordinates, another theorem was presented in reference~\citenum{Lobo:2020a}. Concerning satisfaction/violation of the standard point-wise energy conditions of GR, it states:
\begin{theorem}
    For any static anisotropic fluid sphere with line element as in Eq.~(\ref{Buch}), all of the standard point-wise energy conditions are violated whenever $f(r)\neq 0$, $\Sigma(r)>0$, and $\Sigma''(r)>0$.
\end{theorem}
In the same article~\cite{Lobo:2020a}, further decomposition of Eq.~(\ref{Buch}) \emph{via} $f(r) = 1-2M(r)/\Sigma(r)$ was performed to allow for a two-parameter extension of SV spacetime. One fixes
\begin{equation}\label{FanWang}
    \Sigma(r) = \sqrt{r^2+\ell^2} \ , \qquad M(r) = \frac{mr^k\sqrt{r^2+\ell^2}}{(r^{2n}+\ell^{2n})^{\frac{k+1}{2n}}} \ ,
\end{equation}
with the form for $M(r)$ mathematically inspired by the Fan--Wang mass function for regular black holes~\cite{Fan:2016}. The resulting class of geometries has the following properties:
\begin{itemize}
    \item SV spacetime is recovered \emph{via} fixing $n=1,k=0$;
    \item $\forall \ n,k\in\mathbb{Z}^{+}$, Schwarzschild in the usual curvature coordinates is recovered as $\ell\rightarrow0$;
    \item Enforcing $\vert\ell\vert>0$, $\forall \ n,k\in\mathbb{Z}^{+}$ the candidate geometry is globally regular.
\end{itemize}
Several models of interest possessing vibrant causal structures can be explored using Eq.~(\ref{FanWang}), each corresponding to various fixed values of both $n$ and $k$. All models in some fashion smoothly interpolate between regular black holes and traversable wormholes. Several other model geometries were also explored in reference~\citenum{Lobo:2020a}, \emph{via} certain modifications to the $M(r)$ present in Eq.~(\ref{FanWang}), all within the geometric context of static spherical symmetry in Buchdahl coordinates.


\section{Spherical symmetry with dynamics}

\subsection{Vaidya black-bounce}

\begin{figure}[!htb]
\begin{center}
\includegraphics[scale=0.58]{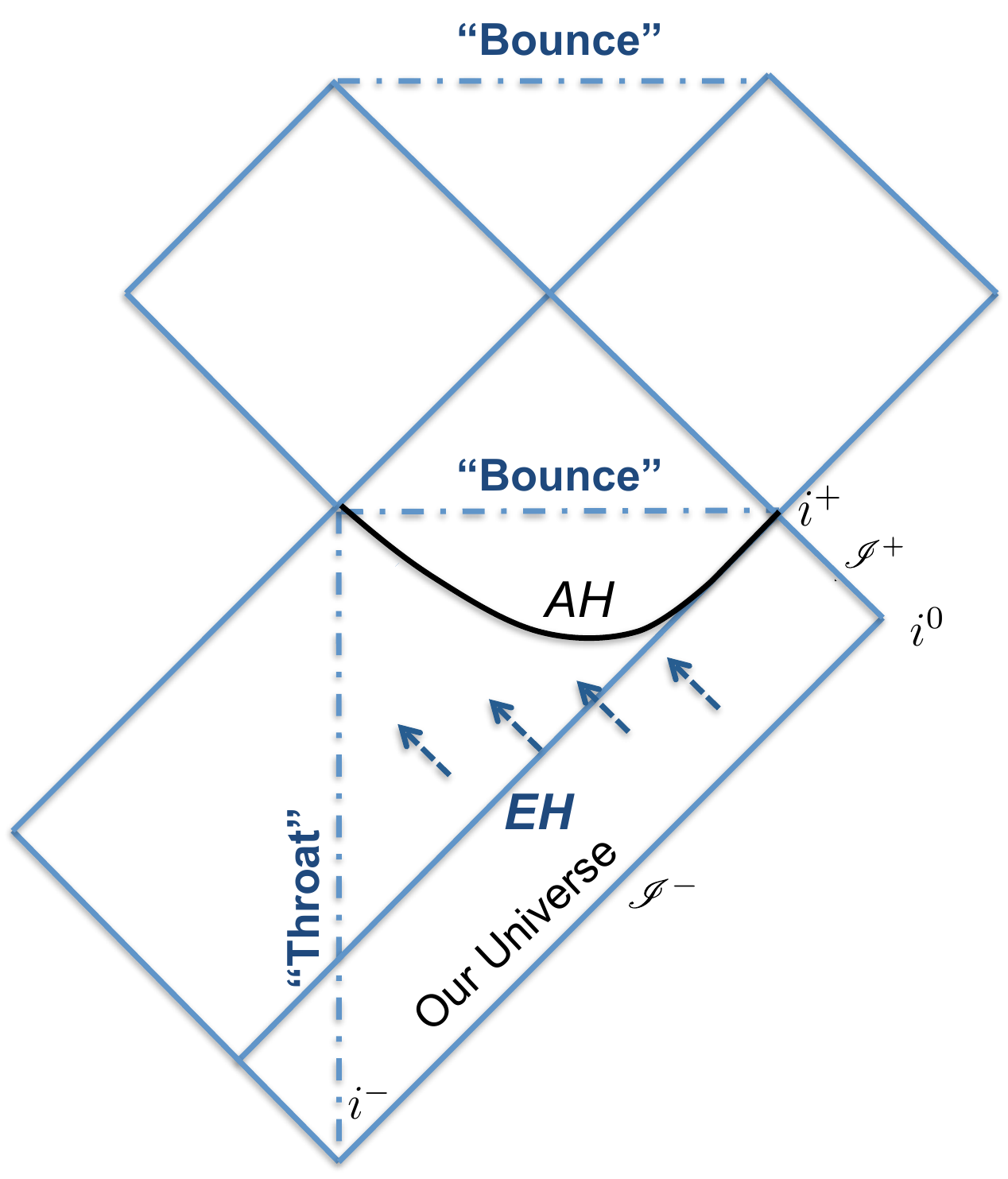}
\end{center}
\caption{Carter--Penrose diagram for a wormhole to black-bounce transition \emph{via} accretion of null dust.}
\label{F:5}
\end{figure}

\begin{figure}[!htb]
\begin{center}
\includegraphics[scale=0.60]{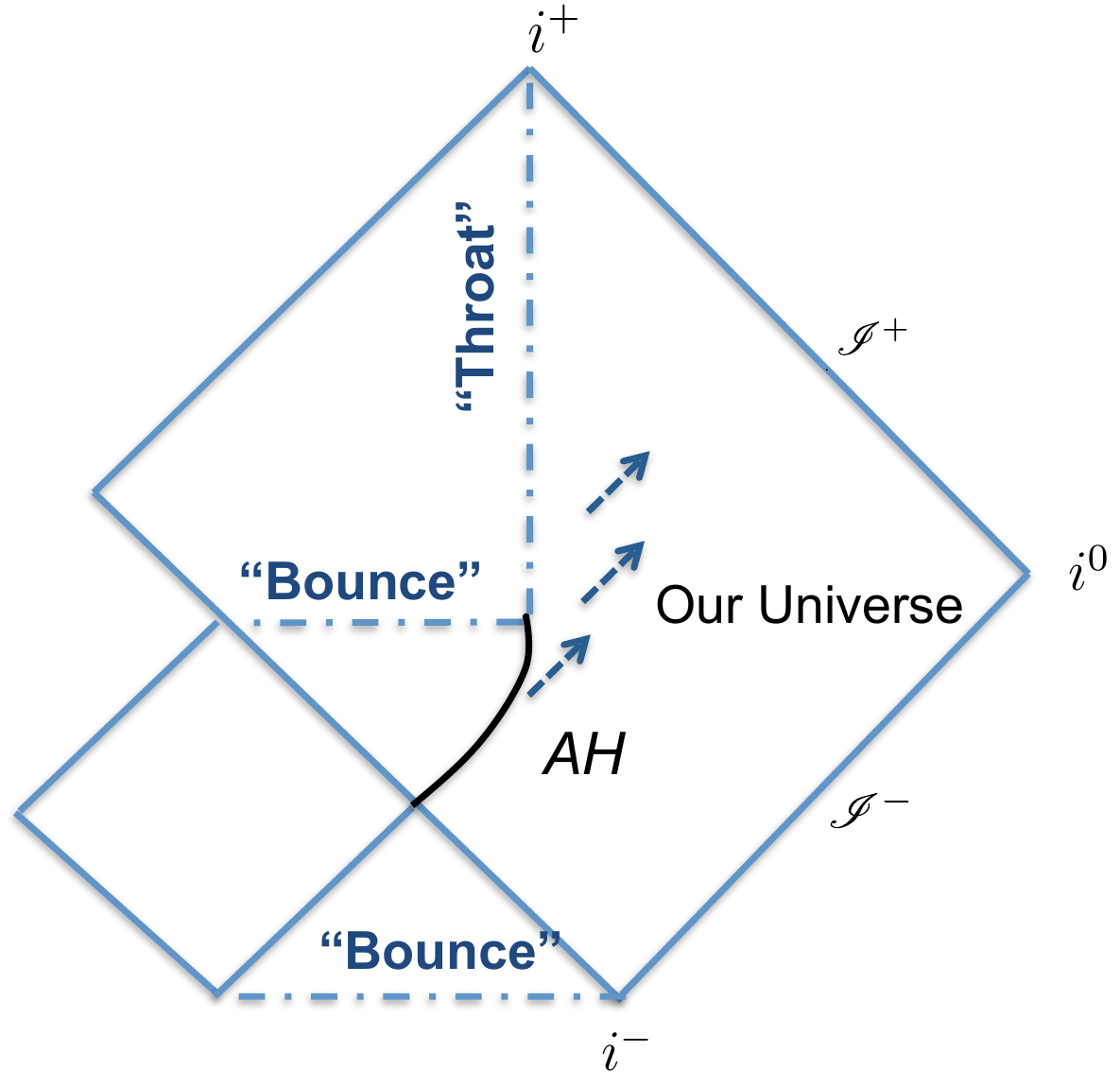}
\qquad
\end{center}
\caption{Carter--Penrose diagram for a black-bounce to wormhole transition due to the emission of positive energy. This is essentially the back reaction of the Hawking evaporation for a semi-classical regular black hole; in this case leaving a wormhole remnant.}
\label{F:6}
\end{figure}

The SV metric was elevated to the regime of dynamical spherical symmetry in reference~\citenum{Simpson:2019}. One first rewrites the line element from Eq.~(\ref{SV}) in Eddington--Finkelstein coordinates, before playing a Vaidya-like `trick' by allowing the mass parameter $m$ to be a function of the null time coordinate $w$. The resulting candidate spacetime is given by the following line element
\begin{equation}
    \dd s^2 = -\left(1-\frac{2m(w)}{\sqrt{r^2+\ell^2}}\right)\,\dd w^2 - (\pm2\,\dd w\dd r) + (r^2+\ell^2)\,\dd\Omega^2_2 \ ,
\end{equation}
where $w=\lbrace u,v\rbrace$ denotes the \emph{outgoing/ingoing} null time coordinate, representing \emph{retarded/advanced} time respectively. In the limit as $m(w)\rightarrow m$, SV spacetime is recovered (in Eddington--Finkelstein coordinates). Introducing dynamics in this specific manner allows for one to discuss simple phenomenological models of an evolving regular black hole/traversable wormhole geometry, either \emph{via} net accretion or net evaporation, whilst still keeping the discourse mathematically tractable. Analysis of the radial null curves yields a dynamical horizon location
\begin{equation}
    r_H(w) = \pm\sqrt{\left[2m(w)\right]^2-\ell^2} \ .
\end{equation}
While this permits analysis of numerous phenomenological models, the most qualitatively interesting are:
\begin{itemize}
    \item Increasing $m(v)$ crossing the $\ell/2$ limit describing the conversion of a traversable wormhole into a regular black hole \emph{via} the accretion of null dust. This qualitative scenario is depicted in Fig.~\ref{F:5};
    \item Decreasing $m(u)$ crossing the $\ell/2$ limit describes the evaporation of a regular black hole leaving a traversable wormhole remnant. The causal structure is depicted in Fig.~\ref{F:6}. This phenomena is causally equivalent to a regular black hole transmuting into a traversable wormhole \emph{via} the accretion of phantom energy (though mathematically this scenario would instead correspond to increasing $m(v)$ crossing the $\ell/2$ limit).
\end{itemize}
%
%
General discussion concerning evaporation/accretion models involving both regular black holes and traversable wormholes can be found in references~\citenum{Babichev:2004a, Babichev:2004b, Martin-Moruno:2006, Gonzalez-Diaz:2007, Martin-Moruno:2007, Madrid:2010, Lobo:2013, Chakrabarti:2021}. It is worth emphasising that one is able to classically describe the transmutation of a wormhole into a regular black hole, or \emph{vice versa}, only because the curvature-regularity of the black hole implies there is no topology change.

In what has become a typical result for nonsingular black hole mimickers in standard GR, the candidate spacetime is found to be in global violation of the NEC. For the `Vaidya black-bounce' spacetime, the existence of the radial null vector $k^\mu=(0,1,0,0)$ implies that one has
\begin{equation}
    T_{\mu\nu}k^{\mu}k^{\nu} \varpropto G_{\mu\nu}k^{\mu}k^{\nu} = -\frac{2\ell^2}{(r^2+\ell^2)^2} < 0 \ .
\end{equation}
For a full analysis of the nonzero curvature tensor components and Riemann curvature invariants for the Vaidya black-bounce geometry, as well as a more detailed discussion surrounding energy condition satisfaction/violation and various phenomenological models, please see reference~\citenum{Simpson:2019}.

\subsection{Thin-shell black-bounce traversable wormhole} 

A thin-shell traversable wormhole variant of SV spacetime was constructed and analysed in reference~\citenum{Lobo:2020b}. The thin-shell surgery is performed \emph{via} a `cut and paste' procedure which has its mathematical roots in the Darmois--Israel formalism~\cite{Israel:1966}. Physically, the geometrodynamics are kept as close to standard GR as possible \emph{via} examination of the Lanczos equations (please see references~\citenum{Visser:1989, Poisson:1995, Visser:1995} for details pertaining to thin-shell surgery).

In the thin-shell construction~\cite{Lobo:2020b}, one analyses both the stability and the evolution of an SV thin-shell traversable wormhole under linearized radial perturbation of the throat. The `bounce' parameter $\ell$ induces a nonzero flux term arising from the conservation identity due to the second contracted Gauss--Codazzi--Mainardi equation. Consequently, enforcing a nonzero $\ell$ leads to an additional constraint on the stability analysis. Several qualitative examples from the parameter space are explored, and the viable stability regions in each case are readily extracted.

\section{Stationary axisymmetry}\label{S:axi}

\subsection{Black-bounce Kerr--Newman}

The first entry from the family of `black-bounce' spacetimes belonging to the geometrical regime of stationary axisymmetry was constructed and explored by Liberati \emph{et al.} in reference~\citenum{Mazza:2021}. This spacetime was constructed \emph{via} application of the Newman--Janis procedure to standard SV spacetime, and acts as the black-bounce analog to Kerr spacetime. Given that in any real, astrophysical setting, one shall have both rotation and the condition that $\vert Q\vert/m\ll 1$, this is the most astrophysically relevant member of the family of black-bounce spacetimes. The black-bounce analog to Kerr--Newman spacetime was then discovered in reference~\citenum{Franzin:2021}; this was obtained \emph{via} application of the regularisation procedure outlined in \S~\ref{Ss:bbRN}, applied to the Kerr--Newman geometry in standard Boyer--Lindquist (BL) coordinates. The line element is given by
\begin{equation}\label{bbKN}
    \dd s^2 = -\frac{\Delta}{\rho^2}(a\sin^2\theta\,\dd\phi-\dd t)^2 + \frac{\sin^2\theta}{\rho^2}[(r^2+\ell^2+a^2)\,\dd\phi-a\,\dd t]^2 + \frac{\rho^2}{\Delta}\,\dd r^2 + \rho^2\,\dd\theta^2 \ ,
\end{equation}
with the usual Kerr quantities modified to be
\begin{equation}
    \rho^2 = r^2+\ell^2+a^2\cos^2\theta \ , \qquad \Delta = r^2+\ell^2+a^2-2m\sqrt{r^2+\ell^2}+Q^2 \ .
\end{equation}
In the astrophysically relevant limit as $Q\rightarrow0$, the Kerr-analog explored in reference~\citenum{Mazza:2021} is recovered precisely. For specifics pertaining to `black-bounce Kerr' (bbK), please consult reference~\citenum{Mazza:2021}. Generally speaking, if one takes the $Q\rightarrow0$ limit for the relevant results obtained for the `black-bounce Kerr--Newman' (bbKN) spacetime (Eq.~(\ref{bbKN})), the analogous result for bbK is recovered precisely. In the limit as the spin parameter $a\rightarrow0$, bbKN reduces to the bbRN geometry from \S~\ref{Ss:bbRN}.

One can see, somewhat trivially, that the natural domains for the angular and temporal coordinates are unaffected when comparing Eq.~(\ref{bbKN}) with standard Kerr--Newman spacetime in BL coordinates. In contrast, the domain for the radial coordinate extends from $r\in[0,+\infty)$ to $r\in(-\infty,+\infty)$, in exactly the same fashion as was seen for bbRN spacetime in \S~\ref{Ss:bbRN}. Examination of the Riemann curvature invariants and orthonormal Riemann tensor components reveals that they are globally finite when $\vert\ell\vert>0$, hence the geometry is free from curvature singularities. Consequently, bbKN qualitatively interpolates between charged rotating regular black holes and charged rotating traversable wormholes. Both the Ricci and Einstein tensors are diagonal in an orthonormal basis.

The causal structure is nontrivial, with horizons characterised by the roots of $\Delta$. This gives the following horizon location(s):
\begin{equation}
    r_{H} = S_{1}\sqrt{(m+S_{2}\sqrt{m^2-Q^2-a^2})^2-\ell^2} \ .
\end{equation}
As it was in \S~\ref{Ss:bbRN}, $S_{1}=S_{2}=\pm1$, with $S_{1}$ fixing which universe one is in, and $S_{2}$ determining whether it is an outer ($S_{2}=+1$) or inner ($S_{2}=-1$) horizon. The qualitatively different possible causal structures are the same as explored in \S~\ref{Ss:bbRN}, with the quantitative aspects slightly shifted due to the presence of the spin parameter $a$. Notably, in the standard Kerr--Newman geometry in BL coordinates one demands $Q^2+a^2\leq m^2$ in order to avoid naked singularities. In the case for bbKN, arbitrary values of both spin and charge may be considered, as in the absence of any horizons one has a traversable wormhole geometry with a rotating throat. Therefore, if one uses bbKN to model astrophysical objects, the weak cosmic censorship conjecture is trivially satisfied.

If horizons are present, their surface gravity is given by
\begin{equation}
    \kappa_{S_{2}}=\frac{1}{2}\frac{\dd}{\dd r}\left(\frac{\Delta}{r^2+\ell^2+a^2}\right)\Bigg\vert_{r=r_{H}} = \kappa_{S_{2}}^{\text{KN}}\sqrt{\frac{r_{H}^2}{r_{H}^2+\ell^2}} \ ,
\end{equation}
where $S_{2}$ fixes whether one is describing the surface gravity of an outer or inner horizon. The ergosurface is characterised by $g_{tt}=0$, giving
\begin{equation}
    r_{\text{erg}} = S_{1}\sqrt{(m+S_{2}\sqrt{m^2-Q^2-a^2\cos^2\theta})^2-\ell^2} \ .
\end{equation}
An informative result pertaining to circular orbits in the equatorial plane for the bbKN spacetime is as follows: if standard Kerr--Newman in BL coordinates has an equatorial circular orbit at $r_{\text{KN},0}$, then bbKN possesses an analogous equatorial circular orbit at coordinate location $r_0=\sqrt{r_{\text{KN},0}^2-\ell^2}$, provided that $r_{\text{KN},0}^2\geq\ell^2$.

The candidate spacetime possesses a nontrivial Killing tensor; upon definition of the objects
\begin{equation}
    l^{\mu} = \left(\frac{r^2+\ell^2+a^2}{\Delta},1,0,\frac{a}{\Delta}\right) \ , \ \ \mbox{and} \ n^{\mu} = \frac{1}{2\rho^2}(r^2+\ell^2+a^2,-\Delta,0,a) \ ,
\end{equation}
one finds that the following is a Killing tensor, \emph{i.e.}, $K_{(\mu\nu;\lambda)}=0$:
\begin{equation}\label{Kill}
    K_{\mu\nu} = \rho^2(l_{\mu}n_{\nu}+l_{\nu}n_{\mu})+(r^2+\ell^2)g_{\mu\nu} \ .
\end{equation}
In this context, the objects $l^{\mu}$ and $n^{\mu}$ are a pair of geodesic null vectors belonging to a generalised Kinnersley tetrad (see reference~\citenum{Mazza:2021} for details). The Killing tensor from Eq.~(\ref{Kill}) has diagonal matrix representation. The existence of a nontrivial Killing tensor directly implies the existence of a generalised Carter constant $\mathcal{C}$, and together with the conserved quantities arising from the temporal and azimuthal Killing vectors (energy and angular momentum per unit mass), as well as the conserved quantity arising from the metric itself (which is a \emph{trivial} Killing tensor), this gives four constants of the motion. This, in principle, means that the geodesics governing the motion of test particles are integrable on the bbKN spacetime. It should be noted that the resulting integral formulae may or may not be analytic.

In reference~\citenum{Baines:2021z}, a refinement was made to Proposition $1.3$ from reference~\citenum{Giorgi:2021}. It should also be noted that this result is implicitly present in reference~\citenum{Benenti:2002}. The result is best summarised by the following theorem:
\begin{theorem}\label{T:4.1}
    Let $(\mathcal{M},g_{\mu\nu})$ be a Lorentzian manifold in possession of a nontrivial Killing tensor $K_{\mu\nu}$. Then upon definition of the Carter operator as: $\mathcal{K}\Phi=\nabla_{\mu}\left(K^{\mu\nu}\nabla_{\nu}\Phi\right)$, and the D'Alembertian scalar wave operator as: $\Box\Phi=\nabla_{\mu}\left(g^{\mu\nu}\nabla_{\nu}\Phi\right)$, there is the following result:
    \begin{equation}
        [\mathcal{K},\Box]\Phi = \frac{2}{3}\left(\nabla_{\mu}[R,K]^{\mu}{}_{\nu}\right)\nabla^{\nu}\Phi \ .
    \end{equation}
    A sufficient condition for this operator commutator to vanish is therefore the commutativity of the Ricci and Killing tensors \emph{via} matrix multiplication; $R^{\mu}{}_{\alpha}K^{\alpha}{}_{\nu}=K^{\mu}{}_{\alpha}R^{\alpha}{}_{\nu}\Longrightarrow[R,K]^{\mu}{}_{\nu}=0$. Hence for any candidate spacetime with a nontrivial Killing tensor, commutativity of the Ricci and Killing tensors \emph{via} matrix multiplication is sufficient to conclude that the massive minimally coupled Klein--Gordon equation is separable on the candidate geometry.
\end{theorem}
In view of Theorem~\ref{T:4.1}, given that both the Ricci and nontrivial Killing tensors for the bbKN geometry have a diagonal matrix representation, one may conclude that the scalar wave equation is separable on bbKN spacetime.

Given the existence of a nontrivial Killing tensor, it is natural to ponder the existence of the full `Killing tower'~\cite{Frolov:2017} in bbKN spacetime. One finds a `would-be' Killing--Yano tensor, of the form
\begin{align}
    f_{\mu \nu} &= \begin{pmatrix}
    0 & -a \cos \theta & 0 & 0 \\
    a \cos \theta & 0 & 0 & - a^2 \cos \theta \sin^2 \theta\\
    0& 0 & 0 & 0\\
    0 & a^2 \cos \theta \sin^2 \theta & 0 & 0
    \end{pmatrix}
\nonumber\\
    &+ \sqrt{r^2+\ell^2} \sin\theta \begin{pmatrix}
    0 & 0 & a  & 0 \\
    0 & 0 & 0 & 0\\
    - a & 0 & 0 & (r^2+\ell^2+a^2)\\
    0 & 0 & - (r^2+\ell^2+a^2) & 0
    \end{pmatrix}.
\end{align}
It is not difficult to verify that $f_{\mu\alpha}f^{\alpha}{}_{\nu}=K_{\mu\nu}$, and as such this is a genuine `square root' of the Killing tensor, however it fails to be a genuine Killing--Yano tensor because $f_{\mu(\nu;\alpha)}\neq0$. It is a known result that a candidate spacetime only possesses a closed conformal Killing--Yano tensor if the spacetime is of Petrov type D~\cite{Frolov:2017}. However, it was proven by Liberati \emph{et al.} in reference~\citenum{Mazza:2021} that the bbK spacetime is Petrov type I, and it is a direct implication of this proof that the bbKN spacetime is also not algebraically special. Consequently, it is an expected result that one can not find a genuine Killing--Yano tensor on bbKN spacetime.

Now coupling the geometry to the Einstein equations of standard GR, the fact that the Einstein tensor is diagonal in an orthonormal basis proves that the stress-energy tensor for bbKN spacetime is Hawking--Ellis type I~\cite{Hawking:1973,Martin-Moruno:2021,Martin-Moruno:2018, Martin-Moruno:2017}. The radial NEC is violated in view of
\begin{equation}
    \varepsilon + p_r = - \frac{\ell^2 \Delta}{{8\pi}\rho^6} \ ,
\end{equation}
where $\varepsilon$ and $p_r$ are energy density and radial pressure respectively.

The stress-energy tensor can be decomposed into gravitational and electromagnetic contributions (similar to bbRN spacetime in \S~\ref{Ss:bbRN}):
\begin{equation}
    \frac{1}{8\pi}G^{\hat{\mu}}{}_{\nu} = T^{\hat{\mu}}{}_{\nu} = \left[T_{\text{bbK}}\right]^{\hat{\mu}}{}_{\hat{\nu}} + \left[T_{Q}\right]^{\hat{\mu}}{}_{\hat{\nu}} \ ,
\end{equation}
where
\begin{equation}
\label{E:xxx}
  [T_Q]^{\hat\mu}{}_{\hat\nu}  = \frac{1}{8\pi} \frac{Q^2 \left(\rho^2-\ell^2\right)}{\rho^6}
  \left[ \mbox{diag}\left(-1,-1,1,1\right) +\frac{2\ell^2}{\rho^2{-\ell^2}} \mbox{diag}\left(1,0,0,0\right)
  \right] \ .
\end{equation}
When comparing Eq.~(\ref{E:xxx}) directly with the charge-dependent contribution to the stress-energy from bbRN spacetime, one simply makes the substitutions
\begin{equation}
    \frac{Q^2 r^2} {(r^2+\ell^2)^3}
    \longleftrightarrow
    \frac{Q^2 \left(\rho^2-\ell^2\right)}{\rho^6}
    \qquad \hbox{and} \qquad
    \frac{2\ell^2}{r^2}
    \longleftrightarrow
    \frac{2\ell^2}{\rho^2{-\ell^2}}.
\end{equation}
Structurally, the first term in Eq.~\eqref{E:xxx} is  of the form of the Maxwell stress-energy tensor. The second term appears to be structurally of the form of charged dust. However, the interpretation of the stress-energy is far less straightforward than the somewhat similar picture present in bbRN spacetime (\S~\ref{Ss:bbRN}). If one performs the substitution $r\rightarrow\sqrt{r^2+\ell^2}$ to the electromagnetic potential from Kerr--Newman spacetime, one finds
\begin{equation}
    A_{\hat \mu} = 
    -\frac{Q \sqrt{r^2+\ell^2}} {\sqrt{\rho^2|\Delta|}} \left( 1,0,0,0 \right) \ ,
    \label{eq:empot2}
\end{equation}
giving the following for the electromagnetic field strength tensor $F_{\hat{\mu}\hat{\nu}}$:
\begin{align}
   F_{\hat t \hat r} = -F_{\hat r \hat t} &=  - \frac{Q}{\rho^4} \sqrt{\frac{r^2 }{r^2+\ell^2}} (r^2 +\ell^2 - a^2 \cos^2\theta)\ ,\\
   F_{\hat \theta \hat \phi} = -F_{\hat \phi \hat \theta} &=  \frac{2aQ\cos\theta \sqrt{r^2+\ell^2}}{\rho^4}\ .
\end{align}
Immediately it is clear that $F_{[\hat{\mu}\hat{\nu},\hat{\sigma}]}=0$. For the inhomogeneous Maxwell equation, defining $z=\sqrt{r^2+\ell^2}$, one has
\begin{align}
    \nabla^{\hat{\mu}} F_{\hat{\mu} \hat{\nu}} &= J_{\hat{\nu}} = \frac{Q\ell^2}{\rho^7 z} \left(-\frac{\Delta \left(\rho^4+2 \rho^2 z^2-4 z^4\right)}{ z^2 \sqrt{|\Delta|}}, 0, 0, {2 a \sin\theta \left(\rho^2-2 z^2\right)}{}\right).\label{eq:maxinhomo}
\end{align}
Interpreting this as an effective electromagnetic source, one has $E_{\hat r} = F_{\hat t \hat r}$ and $B_{\hat r} = F_{\hat\theta \hat\phi}$, giving
\begin{align}
    \left[T_\text{Maxwell} \right]^{\hat\mu}{}_{{\hat\nu}} =  \frac{E_{\hat r}^2 + B_{\hat r}^2}{8\pi} \mbox{diag}\left(-1,-1,1,1\right) \ ,\label{TMaxwell_bbKN}
\end{align}
with
\begin{equation}
    E_{\hat r}^2 +  B_{\hat r}^2
    = \frac{Q^2}{\rho^4} \frac{r^2}{r^2+\ell^2} + \frac{4 Q^2 \ell^2 a^2 \cos^2\theta}{\rho^8} \ .
\end{equation}
Comparison with Eq.~(\ref{E:xxx}) reveals that a more subtle interpretation of the electromagnetic stress-energy than standard Maxwell's is required. Presented herein are two alternatives.

\paragraph{Nonlinear electrodynamics}
One can decompose the electromagnetic stress-energy tensor as follows
\begin{equation}\label{mathcalA}
    [T_Q]^{\hat{\mu}}{}_{\hat{\nu}} = \mathcal{A}\,[T_{\text{Maxwell}}]^{\hat{\mu}}{}_{\hat{\nu}} + \Xi\,V^{\hat{\mu}}V_{\hat{\nu}} \ .
\end{equation}
The $\Xi\,V^{\hat{\mu}}V_{\hat{\nu}}$ term is once again appropriate to interpret as charged dust. Comparing Eq's.~(\ref{TMaxwell_bbKN}) and~(\ref{mathcalA}) allows one to solve for the multiplicative factor $\mathcal{A}$:
\begin{equation}
    \mathcal{A} = \frac{Q^2(\rho^2-\ell^2)}{\rho^6\left(E_{\hat{r}}^2+B_{\hat{r}}^2\right)} = 1 - a^2\ell^2\frac{\cos^2\theta[4(r^2+\ell^2)-\rho^2]}{\rho^4r^2+4a^2\ell^2\cos^2\theta(r^2+\ell^2)} \ .
\end{equation}
It is clear that in both the limit as $\ell\rightarrow0$ (Kerr--Newman), and the limit as $a\rightarrow0$ (bbRN), $\mathcal{A}\rightarrow1$, recovering standard Maxwell's as expected. The fact that $\mathcal{A}$ is independent of the charge $Q$ indicates it is appropriate to interpret this term through the lens of nonlinear electrodynamics (NLED), where generally one has $ \left[T_\text{NLED}\right]^{\hat{\mu}}{}_{\hat{\nu}}\propto  \left[T_\text{Maxwell}\right]^{\hat{\mu}}{}_{\hat{\nu}}$. NLED in the context of regular black holes is prevalent throughout the literature~\cite{Ayon-Beato:2000, Bronnikov:2000, Ayon-Beato:2004, Lobo:2006, Bolokhov:2012, Balart:2014, Bronnikov:2017, Rodrigues:2018}. In the large distance limit one has
\begin{equation}
\mathcal{A} = 1 -\frac{3\ell^2 a^2 \cos^2\theta}{r^4} + \mathcal{O}(r^{-6}) \ ,
\end{equation}
indicating that standard Maxwell's $+$ charged dust is a safe approximation as $r\rightarrow+\infty$. Conversely, for small $r$ one has
\begin{equation}
\mathcal{A} = \frac{\ell^2+a^2\cos\theta^2}{4\ell^2}+ \mathcal{O}(r^2) \ ,
\end{equation}
which allows one to interpret a rescaling of the Maxwell stress-energy deep in the core of the black-bounce. This holistic picture is common in NLED.

\paragraph{Anisotropic fluid}

One can instead ask that
\begin{equation}
\left[T_Q \right]_{\hat \mu \hat \nu} -    \left[T_\text{Maxwell} \right]_{\hat \mu \hat \nu} = \mbox{diag}\left(\varepsilon_{f}, -p_{f}, p_{f}, p_{f}\right),
\label{eq:difference}
\end{equation}
which would imply that (recall $z=\sqrt{r^2+\ell^2}$)
\begin{equation}
    \varepsilon_{f} = \frac{Q^2 \ell^2}{z^2 \rho^8} \left(4z^4-7z^2\rho^2+\rho^4 \right), \quad p_{f} = \frac{Q^2 \ell^2}{z^2 \rho^8} \left(4z^4-5z^2\rho^2+\rho^4 \right).
\end{equation}
Consequently, $[T_Q]^{\hat{\mu}}{}_{\hat{\nu}}$ can be interpreted as standard Maxwell's in the presence of an anisotropic fluid. The anisotropic fluid can then be written as
\begin{equation}
    \varepsilon_{f} V_{\hat \mu} V_{\hat \nu} + \frac{p_{f}}{3} \left(g_{\hat \mu \hat \nu} + V_{\hat \mu} V_{\hat \nu} \right) + \pi_{\hat \mu \hat \nu} \ ,
\end{equation}
with $V^{\hat \mu} = \left(1,0,0,0\right)$ being the velocity of the fluid, and
\begin{equation}
    \pi_{\hat \mu \hat \nu} = \frac{2p_{f}}{3} \mbox{diag}\left(0,-2,1,1\right)
\end{equation}
being the (traceless) anisotropic shear~\cite{Hawking:1973}.

\section{Discussion}

The family of black-bounce spacetimes contains many different candidate geometries which smoothly interpolate between regular black holes and traversable wormholes in standard GR, both with and without rotation, both with and without electrical charge, and both with and without dynamics. The high degree of mathematical tractability of the black-bounce spacetimes means that all models are amenable to the straightforward extraction of astrophysical observables, which \emph{may} be falsifiable/verifiable by the experimental community.

Future research options abound. Given the knowledge of Klein--Gordon separability on bbKN spacetime, one should also test for Maxwell separability on the bbKN spacetime. The knowledge of Klein--Gordon separability already implies that bbKN is amenable to a standard spin zero quasi-normal modes analysis, and if Maxwell's equations also separate then this would extend to a spin one analysis. Further investigation into the spin two axial and polar modes may eventually lead to a measurable ringdown signal able to be captured by LIGO/VIRGO (or LISA) and compared with the analogous signal from Kerr-like mergers. Alternative lines of inquiry for future research could involve attempting to provide a physical mechanism for the bounce, or possibly discovering precisely which mathematical modifications to the Einstein equations would be required such that SV is the unique solution in static spherical symmetry in the new framework.

\section*{Acknowledgements}

AS was supported by a Victoria University of Wellington PhD scholarship, and was also indirectly supported by the Marsden Fund, \emph{via} a grant administered by the Royal Society of New Zealand.\\
AS would also like to acknowledge the following list of authors, all of whom have contributed in meaningful ways towards the development of the family of black-bounce spacetimes: Matt Visser, Prado Mart\'{i}n--Moruno, Francisco S. N. Lobo, Manuel E. Rodrigues, Marcos V. d. S. Silva, Stefano Liberati, Edgardo Franzin, and Jacopo Mazza.




\begin{thebibliography}{99}

\enlargethispage{40pt}
\bibitem{ligo-detection-papers}
See \url{https://www.ligo.caltech.edu/page/detection-companion-papers} for a collection of detection papers from LIGO. \\
See also \url{https://pnp.ligo.org/ppcomm/Papers.html} for a complete list of publications from the LIGO Scientific Collaboration and Virgo Collaboration.


\bibitem{grav-wave-observations-wiki}
See, for example, \href{https://en.wikipedia.org/wiki/List_of_gravitational_wave_observations}{wikipedia.org/List\_of\_gravitational\_wave\_observations} for a list of current (May 2020) gravitational wave observations.


\bibitem{LISA}
E.~Barausse, E.~Berti, T.~Hertog, S.~A.~Hughes, P.~Jetzer, P.~Pani, T.~P.~Sotiriou, N.~Tamanini, H.~Witek and K.~Yagi, \textit{et al.},
``Prospects for Fundamental Physics with LISA'',
Gen. Rel. Grav. \textbf{52} (2020) no.8, 81,
doi:10.1007/s10714-020-02691-1,
[\href{https://arxiv.org/pdf/2001.09793.pdf}{arXiv:2001.09793} [gr-qc]].


\bibitem{Eiroa:2010}
E.~F.~Eiroa and C.~M.~Sendra,
``Gravitational lensing by a regular black hole'',
Class. Quant. Grav. \textbf{28} (2011), 085008,
doi:10.1088/0264-9381/28/8/085008,
[\href{https://arxiv.org/pdf/1011.2455.pdf}{arXiv:1011.2455} [gr-qc]].

\bibitem{Flachi:2012}
A.~Flachi and J.~P.~S.~Lemos,
``Quasinormal modes of regular black holes'',
Phys. Rev. D \textbf{87} (2013) no.2, 024034,
doi:10.1103/PhysRevD.87.024034,
[\href{https://arxiv.org/pdf/1211.6212.pdf}{arXiv:1211.6212} [gr-qc]].

\bibitem{Abdujabbarov:2016}
A.~Abdujabbarov, M.~Amir, B.~Ahmedov and S.~G.~Ghosh,
``Shadow of rotating regular black holes'',
Phys. Rev. D \textbf{93} (2016) no.10, 104004,
doi:10.1103/PhysRevD.93.104004,
[\href{https://arxiv.org/pdf/1604.03809.pdf}{arXiv:1604.03809} [gr-qc]].

\bibitem{Carballo-Rubio:2018}
R.~Carballo-Rubio, F.~Di Filippo, S.~Liberati, C.~Pacilio and M.~Visser,
``On the viability of regular black holes'',
JHEP \textbf{07} (2018), 023,
doi:10.1007/JHEP07(2018)023,
[\href{https://arxiv.org/pdf/1805.02675.pdf}{arXiv:1805.02675} [gr-qc]].

\bibitem{Carballo-Rubio:2019a}
R.~Carballo-Rubio, F.~Di Filippo, S.~Liberati and M.~Visser,
``Opening the Pandora\textquoteright{}s box at the core of black holes'',
Class. Quant. Grav. \textbf{37} (2020) no.14, 14,
doi:10.1088/1361-6382/ab8141,
[\href{https://arxiv.org/pdf/1908.03261.pdf}{arXiv:1908.03261} [gr-qc]].

\bibitem{Carballo-Rubio:2019b}
R.~Carballo-Rubio, F.~Di Filippo, S.~Liberati and M.~Visser,
``Geodesically complete black holes'',
Phys. Rev. D \textbf{101} (2020), 084047,
doi:10.1103/PhysRevD.101.084047,
[\href{https://arxiv.org/pdf/1911.11200.pdf}{arXiv:1911.11200} [gr-qc]].

\bibitem{Dai:2019}
D.~C.~Dai and D.~Stojkovic,
``Observing a Wormhole'',
Phys. Rev. D \textbf{100} (2019) no.8, 083513,
doi:10.1103/PhysRevD.100.083513,
[\href{https://arxiv.org/pdf/1910.00429.pdf}{arXiv:1910.00429} [gr-qc]].

\bibitem{Simonetti:2020}
J.~H.~Simonetti, M.~J.~Kavic, D.~Minic, D.~Stojkovic and D.~C.~Dai,
``Sensitive searches for wormholes'',
Phys. Rev. D \textbf{104} (2021) no.8, L081502,
doi:10.1103/PhysRevD.104.L081502,
[\href{https://arxiv.org/pdf/2007.12184.pdf}{arXiv:2007.12184} [gr-qc]].

\bibitem{Berry:2020}
T.~Berry, A.~Simpson and M.~Visser,
``Photon spheres, ISCOs, and OSCOs: Astrophysical observables for regular black holes with asymptotically Minkowski cores'',
Universe \textbf{7} (2020) no.1, 2,
doi:10.3390/universe7010002,
[\href{https://arxiv.org/pdf/2008.13308.pdf}{arXiv:2008.13308} [gr-qc]].

\bibitem{Carballo-Rubio:2021}
R.~Carballo-Rubio, F.~Di Filippo and S.~Liberati,
``Hearts of Darkness: the inside out probing of black holes'', IJMPD (in press),
[\href{https://arxiv.org/pdf/2106.01530.pdf}{arXiv:2106.01530} [gr-qc]].

\bibitem{Bronnikov:2021}
K.~A.~Bronnikov, R.~A.~Konoplya and T.~D.~Pappas,
``General parametrization of wormhole spacetimes and its application to shadows and quasinormal modes'',
Phys. Rev. D \textbf{103} (2021) no.12, 124062,
doi:10.1103/PhysRevD.103.124062,
[\href{https://arxiv.org/pdf/2102.10679.pdf}{arXiv:2102.10679} [gr-qc]].

\bibitem{Churilova:2021}
M.~S.~Churilova, R.~A.~Konoplya, Z.~Stuchlik and A.~Zhidenko,
``Wormholes without exotic matter: quasinormal modes, echoes and shadows'',
[\href{https://arxiv.org/pdf/2107.05977.pdf}{arXiv:2107.05977} [gr-qc]].

\bibitem{Bambi:2021}
C.~Bambi and D.~Stojkovic,
``Astrophysical Wormholes'',
Universe \textbf{7} (2021) no.5, 136,
doi:10.3390/universe7050136,
[\href{https://arxiv.org/pdf/2105.00881.pdf}{arXiv:2105.00881} [gr-qc]].

\bibitem{Simpson:2021biv}
A.~Simpson,
``Ringing of the regular black hole with asymptotically Minkowski core'',
[\href{https://arxiv.org/pdf/2109.11878.pdf}{arXiv:2109.11878} [gr-qc]].


\bibitem{Simpson:2018}
A.~Simpson and M.~Visser,
``Black-bounce to traversable wormhole'',
JCAP \textbf{02} (2019), 042,
doi:10.1088/1475-7516/2019/02/042
[\href{https://arxiv.org/pdf/1812.07114.pdf}{arXiv:1812.07114} [gr-qc]].

\bibitem{Lobo:2020a}
F.~S.~N.~Lobo, M.~E.~Rodrigues, M.~V.~d.~S.~Silva, A.~Simpson and M.~Visser,
``Novel black-bounce spacetimes: wormholes, regularity, energy conditions, and causal structure'',
Phys. Rev. D \textbf{103} (2021) no.8, 084052,
doi:10.1103/PhysRevD.103.084052,
[\href{https://arxiv.org/pdf/2009.12057.pdf}{arXiv:2009.12057} [gr-qc]].

\enlargethispage{40pt}
\bibitem{Simpson:2019}
A.~Simpson, P.~Mart\'{i}n-Moruno and M.~Visser,
``Vaidya spacetimes, black-bounces, and traversable wormholes'',
Class. Quant. Grav. \textbf{36} (2019) no.14, 145007,
doi:10.1088/1361-6382/ab28a5,
[\href{https://arxiv.org/pdf/1902.04232.pdf}{arXiv:1902.04232} [gr-qc]].

\bibitem{Lobo:2020b}
F.~S.~N.~Lobo, A.~Simpson and M.~Visser,
``Dynamic thin-shell black-bounce traversable wormholes'',
Phys. Rev. D \textbf{101} (2020) no.12, 124035,
doi:10.1103/PhysRevD.101.124035,
[\href{https://arxiv.org/pdf/2003.09419.pdf}{arXiv:2003.09419} [gr-qc]].

\bibitem{Mazza:2021}
J.~Mazza, E.~Franzin and S.~Liberati,
``A novel family of rotating black hole mimickers'',
JCAP \textbf{04} (2021), 082,
doi:10.1088/1475-7516/2021/04/082,
[\href{https://arxiv.org/pdf/2102.01105.pdf}{arXiv:2102.01105} [gr-qc]].

\bibitem{Franzin:2021}
E.~Franzin, S.~Liberati, J.~Mazza, A.~Simpson and M.~Visser,
``Charged black-bounce spacetimes'',
JCAP \textbf{07} (2021), 036,
doi:10.1088/1475-7516/2021/07/036,
[\href{https://arxiv.org/pdf/2104.11376.pdf}{arXiv:2104.11376} [gr-qc]].


\bibitem{Bardeen:1968}
    J.~M.~Bardeen, ``Non-singular general relativistic gravitational collapse'',
    Abstracts of the 5th international conference on gravitation and the theory of relativity (GR5), 
    eds. V.~A.~Fock \emph{et al.} (Tbilisi University Press, Tblisi, Georgia,  former USSR, 1968), 
    pages 174–-175. 
    
\bibitem{Ayon-Beato:2000}
E.~Ayon-Beato and A.~Garcia,
``The Bardeen model as a nonlinear magnetic monopole'',
Phys. Lett. B \textbf{493} (2000), 149--152,
doi:10.1016/S0370-2693(00)01125-4,
[\href{https://arxiv.org/pdf/gr-qc/0009077.pdf}{arXiv:gr-qc/0009077} [gr-qc]].

\bibitem{Hayward:2005}
S.~A.~Hayward,
``Formation and evaporation of regular black holes'',
Phys. Rev. Lett. \textbf{96} (2006), 031103,
doi:10.1103/PhysRevLett.96.031103,
[\href{https://arxiv.org/pdf/gr-qc/0506126.pdf}{arXiv:gr-qc/0506126} [gr-qc]].
    
\bibitem{Bambi:2013}
C.~Bambi and L.~Modesto,
``Rotating regular black holes'',
Phys. Lett. B \textbf{721} (2013), 329--334,
doi:10.1016/j.physletb.2013.03.025,
[\href{https://arxiv.org/pdf/1302.6075.pdf}{arXiv:1302.6075} [gr-qc]].

\bibitem{Frolov:2014}
V.~P.~Frolov,
``Information loss problem and a 'black hole` model with a closed apparent horizon'',
JHEP \textbf{05} (2014), 049,
doi:10.1007/JHEP05(2014)049,
[\href{https://arxiv.org/pdf/1402.5446.pdf}{arXiv:1402.5446} [hep-th]].
    
\bibitem{Neves:2014}
J.~C.~S.~Neves and A.~Saa,
``Regular rotating black holes and the weak energy condition'',
Phys. Lett. B \textbf{734} (2014), 44--48,
doi:10.1016/j.physletb.2014.05.026,
[\href{https://arxiv.org/pdf/1402.2694.pdf}{arXiv:1402.2694} [gr-qc]].
    
\bibitem{Fan:2016}
Z.~Y.~Fan and X.~Wang,
``Construction of Regular Black Holes in General Relativity'',
Phys. Rev. D \textbf{94} (2016) no.12, 124027,
doi:10.1103/PhysRevD.94.124027,
[\href{https://arxiv.org/pdf/1610.02636.pdf}{arXiv:1610.02636} [gr-qc]].


\bibitem{Morris:1988a}
M.~S.~Morris and K.~S.~Thorne,
``Wormholes in space-time and their use for interstellar travel: A tool for teaching general relativity'',
Am. J. Phys. \textbf{56} (1988), 395--412,
doi:10.1119/1.15620.

\bibitem{Morris:1988b}
M.~S.~Morris, K.~S.~Thorne and U.~Yurtsever,
``Wormholes, Time Machines, and the Weak Energy Condition'',
Phys. Rev. Lett. \textbf{61} (1988), 1446--1449,
doi:10.1103/PhysRevLett.61.1446.

\bibitem{Visser:1989a}
M.~Visser,
``Traversable wormholes: Some simple examples'',
Phys. Rev. D \textbf{39} (1989), 3182--3184,
doi:10.1103/PhysRevD.39.3182,
[\href{https://arxiv.org/pdf/0809.0907.pdf}{arXiv:0809.0907} [gr-qc]].

\bibitem{Visser:1989b}
M.~Visser,
``Traversable wormholes from surgically modified Schwarzschild space-times'',
Nucl. Phys. B \textbf{328} (1989), 203--212,
doi:10.1016/0550-3213(89)90100-4,
[\href{https://arxiv.org/pdf/0809.0927.pdf}{arXiv:0809.0927} [gr-qc]].


\bibitem{SVinspirepage}
See \href{https://inspirehep.net/literature?sort=mostrecent\&size=25\&page=1\&q=refersto\%3Arecid\%3A1709812\&ui-citation-summary=true}{inspirehep.net} for a full list of pre-published and published papers citing reference~\cite{Simpson:2018}.


\bibitem{Bronnikov:2019}
K.~A.~Bronnikov and R.~A.~Konoplya,
``Echoes in brane worlds: ringing at a black hole--wormhole transition'',
Phys. Rev. D \textbf{101} (2020) no.6, 064004,
doi:10.1103/PhysRevD.101.064004,
[\href{https://arxiv.org/pdf/1912.05315.pdf}{arXiv:1912.05315} [gr-qc]].

\bibitem{Churilova:2019}
M.~S.~Churilova and Z.~Stuchlik,
``Ringing of the regular black-hole/wormhole transition'',
Class. Quant. Grav. \textbf{37} (2020) no.7, 075014,
doi:10.1088/1361-6382/ab7717,
[\href{https://arxiv.org/pdf/1911.11823.pdf}{arXiv:1911.11823} [gr-qc]].

\bibitem{Islam:2021}
S.~U.~Islam, J.~Kumar and S.~G.~Ghosh,
``Strong gravitational lensing by rotating Simpson--Visser black holes'',
[\href{https://arxiv.org/pdf/2104.00696.pdf}{arXiv:2104.00696} [gr-qc]].

\enlargethispage{40pt}
\bibitem{Tsukamoto:2021}
N.~Tsukamoto,
``Gravitational lensing by two photon spheres in a black-bounce spacetime in strong deflection limits'',
Phys. Rev. D \textbf{104} (2021) no.6, 064022,
doi:10.1103/PhysRevD.104.064022,
[\href{https://arxiv.org/pdf/2105.14336.pdf}{arXiv:2105.14336} [gr-qc]].

\bibitem{Guerrero:2021}
M.~Guerrero, G.~J.~Olmo, D.~Rubiera-Garcia and D.~S.~C.~G\'omez,
``Shadows and optical appearance of black bounces illuminated by a thin accretion disk'',
JCAP \textbf{08} (2021), 036,
doi:10.1088/1475-7516/2021/08/036,
[\href{https://arxiv.org/pdf/2105.15073.pdf}{arXiv:2105.15073} [gr-qc]].

\bibitem{Lima:2021}
H.~C.~D.~Lima, Junior., L.~C.~B.~Crispino, P.~V.~P.~Cunha and C.~A.~R.~Herdeiro,
``Can different black holes cast the same shadow?'',
Phys. Rev. D \textbf{103} (2021) no.8, 084040,
doi:10.1103/PhysRevD.103.084040,
[\href{https://arxiv.org/pdf/2102.07034.pdf}{arXiv:2102.07034} [gr-qc]].

\bibitem{Tsukamoto:2020}
N.~Tsukamoto,
``Gravitational lensing in the Simpson-Visser black-bounce spacetime in a strong deflection limit'',
Phys. Rev. D \textbf{103} (2021) no.2, 024033,
doi:10.1103/PhysRevD.103.024033,
[\href{https://arxiv.org/pdf/2011.03932.pdf}{arXiv:2011.03932} [gr-qc]].

\bibitem{Nascimento:2020}
J.~R.~Nascimento, A.~Y.~Petrov, P.~J.~Porfirio and A.~R.~Soares,
``Gravitational lensing in black-bounce spacetimes'',
Phys. Rev. D \textbf{102} (2020) no.4, 044021,
doi:10.1103/PhysRevD.102.044021,
[\href{https://arxiv.org/pdf/2005.13096.pdf}{arXiv:2005.13096} [gr-qc]].

\bibitem{Zhou:2020}
T.~Y.~Zhou and Y.~Xie,
``Precessing and periodic motions around a black-bounce/traversable wormhole'',
Eur. Phys. J. C \textbf{80} (2020) no.11, 1070,
doi:10.1140/epjc/s10052-020-08661-w.


\bibitem{Delgaty:1998}
M.~S.~R.~Delgaty and K.~Lake,
``Physical acceptability of isolated, static, spherically symmetric, perfect fluid solutions of Einstein's equations'',
Comput. Phys. Commun. \textbf{115} (1998), 395--415,
doi:10.1016/S0010-4655(98)00130-1,
[\href{https://arxiv.org/pdf/gr-qc/9809013.pdf}{arXiv:gr-qc/9809013} [gr-qc]].

\bibitem{Finch:1998}
M.~R.~Finch and J.~E.~F.~Skea, ``A review of the relativistic static fluid sphere", 1998, unpublished.

\bibitem{Boonserm:2007a}
P.~Boonserm and M.~Visser,
``Buchdahl-like transformations for perfect fluid spheres'',
Int. J. Mod. Phys. D \textbf{17} (2008), 135--163,
doi:10.1142/S0218271808011912,
[\href{https://arxiv.org/pdf/0707.0146.pdf}{arXiv:0707.0146} [gr-qc]].

\bibitem{Boonserm:2007b}
P.~Boonserm and M.~Visser, ``Buchdahl-like transformations in general relativity", Thai Journal of Mathematics \textbf{5} \#\textbf{2} (2007) 209--223.

\bibitem{Semiz:2020}
\.I.~Semiz,
``On the (non)genericity of the Kiselev spacetime'',
doi:10.1088/2633-1357/aba1f5,
[\href{https://arxiv.org/pdf/2001.06310.pdf}{arXiv:2001.06310} [gr-qc]].


\bibitem{Babichev:2004a}
E.~Babichev, V.~Dokuchaev and Y.~Eroshenko,
``Black hole mass decreasing due to phantom energy accretion'',
Phys. Rev. Lett. \textbf{93} (2004), 021102,
doi:10.1103/PhysRevLett.93.021102,
[\href{https://arxiv.org/pdf/gr-qc/0402089.pdf}{arXiv:gr-qc/0402089} [gr-qc]].

\bibitem{Babichev:2004b}
E.~Babichev, V.~Dokuchaev and Y.~Eroshenko,
``Dark energy cosmology with generalized linear equation of state'',
Class. Quant. Grav. \textbf{22} (2005), 143--154,
doi:10.1088/0264-9381/22/1/010,
[\href{https://arxiv.org/pdf/astro-ph/0407190.pdf}{arXiv:astro-ph/0407190} [astro-ph]].

\bibitem{Martin-Moruno:2006}
P.~Mart\'{i}n-Moruno, J.~A.~J.~Madrid and P.~F.~Gonz\'{a}lez-D\'{i}az,
``Will black holes eventually engulf the universe?'',
Phys. Lett. B \textbf{640} (2006), 117--120,
doi:10.1016/j.physletb.2006.07.067,
[\href{https://arxiv.org/pdf/astro-ph/0603761.pdf}{arXiv:astro-ph/0603761} [astro-ph]].

\bibitem{Gonzalez-Diaz:2007}
P.~F.~Gonz\'{a}lez-D\'{i}az and P.~Mart\'{i}n-Moruno,
``Wormholes in the accelerating universe'',
doi:10.1142/9789812834300\_0358,
[\href{https://arxiv.org/pdf/0704.1731.pdf}{arXiv:0704.1731} [astro-ph]].

\bibitem{Martin-Moruno:2007}
P.~Mart\'{i}n-Moruno,
``On the formalism of dark energy accretion onto black- and worm-holes'',
Phys. Lett. B \textbf{659} (2008), 40--44,
doi:10.1016/j.physletb.2007.10.083,
[\href{https://arxiv.org/pdf/0709.4410.pdf}{arXiv:0709.4410} [astro-ph]].

\bibitem{Madrid:2010}
J.~A.~J.~Madrid and P.~Mart\'{i}n-Moruno,
``On accretion of dark energy onto black- and worm-holes'', (2010),
[\href{https://arxiv.org/pdf/1004.1428.pdf}{arXiv:1004.1428} [astro-ph.CO]].

\bibitem{Lobo:2013}
F.~S.~N.~Lobo, J.~Martinez-Asencio, G.~J.~Olmo and D.~Rubiera-Garcia,
``Planck scale physics and topology change through an exactly solvable model'',
Phys. Lett. B \textbf{731} (2014), 163--167,
doi:10.1016/j.physletb.2014.02.038,
[\href{https://arxiv.org/pdf/1311.5712.pdf}{arXiv:1311.5712} [hep-th]].


\bibitem{Chakrabarti:2021}
S.~Chakrabarti and S.~Kar,
``Wormhole geometry from gravitational collapse'',
Phys. Rev. D \textbf{104} (2021) no.2, 024071,
doi:10.1103/PhysRevD.104.024071,
[\href{https://arxiv.org/pdf/2106.14761.pdf}{arXiv:2106.14761} [gr-qc]].

\enlargethispage{40pt}
\bibitem{Israel:1966}
W.~Israel,
``Singular hypersurfaces and thin shells in general relativity'',
Nuovo Cim. B \textbf{44S10} (1966), 1
[erratum: Nuovo Cim. B \textbf{48} (1967), 463],
doi:10.1007/BF02710419

\bibitem{Visser:1989}
M.~Visser,
``Traversable wormholes from surgically modified Schwarzschild space-times'',
Nucl. Phys. B \textbf{328} (1989), 203--212,
doi:10.1016/0550-3213(89)90100-4,
[\href{https://arxiv.org/pdf/0809.0927.pdf}{arXiv:0809.0927} [gr-qc]].

\bibitem{Poisson:1995}
E.~Poisson and M.~Visser,
``Thin shell wormholes: Linearization stability'',
Phys. Rev. D \textbf{52} (1995), 7318--7321,
doi:10.1103/PhysRevD.52.7318,
[\href{https://arxiv.org/pdf/gr-qc/9506083.pdf}{arXiv:gr-qc/9506083} [gr-qc]].

\bibitem{Visser:1995}
M.~Visser,
``Lorentzian wormholes: From Einstein to Hawking''.


\bibitem{Baines:2021z}
J.~Baines, T.~Berry, A.~Simpson and M.~Visser,
``Killing tensor and Carter constant for Painleve-Gullstrand form of Lense-Thirring spacetime'',
[\href{https://arxiv.org/pdf/2110.01814.pdf}{arXiv:2110.01814} [gr-qc]].

\bibitem{Giorgi:2021}
E.~Giorgi,
``The Carter tensor and the physical-space analysis in perturbations of Kerr-Newman spacetime'',
[\href{https://arxiv.org/pdf/2105.14379.pdf}{arXiv:2105.14379} [gr-qc]].

\bibitem{Benenti:2002}
S.~Benenti, C.~Chanu, and G.~Rastelli, ``Remarks on the connection between the additive separation of the Hamilton--Jacobi equation and the multiplicative separation of the Schr{\"o}dinger equation. I. The completeness and Robertson conditions", Journal of Mathematical Physics, \textbf{43} (2002) no.11, 5183--5222, American Institute of Physics,
doi:10.1063/1.1506180 .


\bibitem{Frolov:2017}
V.~Frolov, P.~Krtous and D.~Kubiznak,
``Black holes, hidden symmetries, and complete integrability'',
Living Rev. Rel. \textbf{20} (2017) no.1, 6,
doi:10.1007/s41114-017-0009-9,
[\href{https://arxiv.org/pdf/1705.05482.pdf}{arXiv:1705.05482} [gr-qc]].

\bibitem{Hawking:1973}
S.~W.~Hawking and G.~F.~R.~Ellis,
``The Large Scale Structure of Space-Time'',
doi:10.1017/CBO9780511524646.

\bibitem{Martin-Moruno:2017}
P.~Mart\'\i{}n-Moruno and M.~Visser,
``Generalized Rainich conditions, generalized stress-energy conditions, and the Hawking-Ellis classification'',
Class. Quant. Grav. \textbf{34} (2017) no.22, 225014,
doi:10.1088/1361-6382/aa9039,
[\href{https://arxiv.org/pdf/1707.04172.pdf}{arXiv:1707.04172} [gr-qc]].

\bibitem{Martin-Moruno:2018}
P.~Mart\'{i}n-Moruno and M.~Visser,
``Essential core of the Hawking\textendash{}Ellis types'',
Class. Quant. Grav. \textbf{35} (2018) no.12, 125003,
doi:10.1088/1361-6382/aac147,
[\href{https://arxiv.org/pdf/1802.00865}{arXiv:1802.00865} [gr-qc]].

\bibitem{Martin-Moruno:2021}
P.~Mart\'{i}n-Moruno and M.~Visser,
``Hawking-Ellis classification of stress-energy tensors: Test fields versus backreaction'',
Phys. Rev. D \textbf{103} (2021) no.12, 124003,
doi:10.1103/PhysRevD.103.124003,
[\href{https://arxiv.org/pdf/2102.13551.pdf}{arXiv:2102.13551} [gr-qc]].


\bibitem{Bronnikov:2000}
K.~A.~Bronnikov,
``Regular magnetic black holes and monopoles from nonlinear electrodynamics'',
Phys. Rev. D \textbf{63} (2001), 044005,
doi:10.1103/PhysRevD.63.044005,
[\href{https://arxiv.org/pdf/gr-qc/0006014.pdf}{arXiv:gr-qc/0006014} [gr-qc]].

\bibitem{Ayon-Beato:2004}
E.~Ayon-Beato and A.~Garcia,
``Four parametric regular black hole solution'',
Gen. Rel. Grav. \textbf{37} (2005), 635,
doi:10.1007/s10714-005-0050-y,
[\href{https://arxiv.org/pdf/hep-th/0403229.pdf}{arXiv:hep-th/0403229} [hep-th]].

\bibitem{Lobo:2006}
F.~S.~N.~Lobo and A.~V.~B.~Arellano,
``Gravastars supported by nonlinear electrodynamics'',
Class. Quant. Grav. \textbf{24} (2007), 1069--1088,
doi:10.1088/0264-9381/24/5/004,
[\href{https://arxiv.org/pdf/gr-qc/0611083.pdf}{arXiv:gr-qc/0611083} [gr-qc]].

\bibitem{Bolokhov:2012}
S.~V.~Bolokhov, K.~A.~Bronnikov and M.~V.~Skvortsova,
``Magnetic black universes and wormholes with a phantom scalar'',
Class. Quant. Grav. \textbf{29} (2012), 245006,
doi:10.1088/0264-9381/29/24/245006,
[\href{https://arxiv.org/pdf/1208.4619.pdf}{arXiv:1208.4619} [gr-qc]].

\bibitem{Balart:2014}
L.~Balart and E.~C.~Vagenas,
``Regular black holes with a nonlinear electrodynamics source'',
Phys. Rev. D \textbf{90} (2014) no.12, 124045,
doi:10.1103/PhysRevD.90.124045,
[\href{https://arxiv.org/pdf/1408.0306.pdf}{arXiv:1408.0306} [gr-qc]].

\bibitem{Bronnikov:2017}
K.~A.~Bronnikov,
``Nonlinear electrodynamics, regular black holes and wormholes'',
Int. J. Mod. Phys. D \textbf{27} (2018) no.06, 1841005,
doi:10.1142/S0218271818410055,
[\href{https://arxiv.org/pdf/1711.00087.pdf}{arXiv:1711.00087} [gr-qc]].

\bibitem{Rodrigues:2018}
M.~E.~Rodrigues and M.~V.~d.~Silva,
``Bardeen Regular Black Hole With an Electric Source'',
JCAP \textbf{06} (2018), 025,
doi:10.1088/1475-7516/2018/06/025,
[\href{https://arxiv.org/pdf/1802.05095.pdf}{arXiv:1802.05095} [gr-qc]].

\end{thebibliography}
\end{document}